\documentclass[twocolumn,showpacs,preprintnumbers,prl]{revtex4}
\usepackage{dcolumn}
\usepackage{amssymb}
\usepackage{bm}
\usepackage{graphicx}
\usepackage{float}
\newcommand{\avg}[1]{\left< #1 \right>} 
\newcommand{\abs}[1]{\left| #1 \right|} 
\newcommand{\ket}[1]{\left| #1 \right>} 
\newcommand{\bra}[1]{\left< #1 \right|} 
\newcommand{\braket}[2]{\left< #1 \vphantom{#2} \right| \left. #2 \vphantom{#1} \right>}

\begin{document}

\title{Maximally Entangled Mode, Metal-Insulator Transition and Violation of Entanglement Area Law
in Non-interacting Fermion Ground States
}
\author{Mohammad Pouranvari and Kun Yang}
\affiliation{National High Magnetic Field Laboratory and Department of Physics,
Florida State University, Tallahassee, Florida 32306, USA}
\date{\today}
\begin{abstract}

We study in this work the ground state entanglement properties of two models of non-interacting fermions moving in one-dimension (1D), that exhibit metal-insulator transitions. We find that entanglement entropy grows either logarithmically or in a power-law fashion in the metallic phase, thus violating the (1D version of) entanglement area law. No such violation is found in the insulating phase. We further find that characteristics of {\em single fermion} states at the Fermi energy (which can {\em not} be obtained from the many-fermion Slater determinant) is captured by the lowest energy single fermion mode of the {\em entanglement} Hamiltonian; this is particularly true at the metal-insulator transition point. Our results suggest entanglement is a powerful way to detect metal-insulator transitions, {\em without} knowledge of the Hamiltonian of the system.

\end{abstract}

\maketitle

{\em Introduction} --- Characterizing phases and phase transitions in many-particle systems is the central theme of condensed matter physics. Recently attention has been focused on zero temperature, or quantum, phases and phase transitions. In many cases this can be achieved by scrutinizing the ground state wave function(s), without detailed knowledge of the underlying Hamiltonian, other than reasonable assumptions like locality etc. Entanglement has proven a very useful diagnostic in this endeavor, although perhaps not indispensable. A notable exception to this is metal-insulator transition, which appears to require knowledge about the Hamiltonian (or dynamics) {\em in addition to the many-particle ground state wave function}. We illustrate this by considering metal-insulator transition of non-interacting fermions moving in a random potential. One normally starts by solving the
single-particle Hamiltonian to obtain its eigenvalues and eigenstates, and the many-body ground state is a single Slater determinant obtained by filling single-particle states below the Fermi energy. The system is a metal if the states at the Fermi energy are extended, and an insulator otherwise. On the other hand if one were given the single Slater determinant $|\psi\rangle$, it is {\em not} clear how to determine whether it corresponds a metal or an insulator. The reason is one can {\em not} extract the {\em single particle} eigenstates from the {\em many-particle} Slater determinant $|\psi\rangle$, as any SU($N_F$) ($N_F$ being the number of fermions) transformation within the occupied subspace of single particle states leaves $|\psi\rangle$ invariant. Even if one were given the single particle eigenstates, without knowledge about their energies one
would not know which one is at the Fermi energy, and thus still unable to determine which phase the system is in. The situation is far worse in the presence of interactions; here one does not even have the notion of single particle eigenstates at the Fermi energy. It appears the only way to determine whether the ground state is metallic or insulating is to calculate {\em dynamical} correlation functions like conductivity. It is in this sense that metal-insulator transition is very different from many other quantum phase
transitions, and is intrinsically a {\em dynamical} phase transition.

The purpose of this work is to argue that by inspecting entanglement properties of the ground state {\em alone}, one can determine the metallic or insulating nature of the system, without detailed knowledge of the Hamiltonian or dynamics. Our conclusion is based on detailed numerical studies of two one-dimension (1D) models that exhibit metal-insulator transitions, namely random dimer model (RDM)\cite{dunalp} and power-law random banded model (PRBM)\cite{prbm}. We find the following in both models. (i) The block entanglement entropy (EE) diverges with subsystem size in the metallic phase or at the metal-insulator critical point and thus violates the (1D version of) entanglement area law,
while it saturates in the insulating phase. (ii) The inverse-participation ratio (IPR, see below for definition) of the lowest energy single fermion mode of the {\em entanglement} Hamiltonian (which gives the most contributions to EE; referred to as maximally entangled mode or MEM from now on) has qualitatively similar dependence on parameters to that of the single fermion state at Fermi energy in one case, and the two are essentially the same in the other. (iii)
These two modes have maximum overlaps at the metal-insulator critical point. In particular, points (ii) and (iii) indicate that the state at the Fermi level is well characterized by the MEM, especially at the critical point. This can be viewed as a highly non-trivial extension of the result of Li and Haldane\cite{lihaldane}, who found that the entanglement Hamiltonian resembles the edge Hamiltonian of a system in a topological phase. Our results suggest that the entanglement Hamiltonian captures some aspects of the {\em bulk} Hamiltonian of a non-interacting fermion system near metal-insulator transition. Implication of our results for interacting systems will also be discussed.

\emph{Methods} --- For a system in a pure state $\ket{\psi}$, the density matrix is $\rho= \ket{\psi} \bra{\psi}$. Dividing the system into two subsystems $A$ and $B$ in real space, reduced density
matrix of each subsystem is obtained by tracing over degrees of freedom of the other subsystem: $\rho^{A/B}=tr_{B/A} (\rho)$. Block EE between the two subsystems is $EE=-tr(\rho^{A}\ln{\rho^{A}})=-tr(\rho^{B}\ln{\rho^{B}})$. For a single Slater-determinant ground state,
\begin{equation} \label{rho}
\rho^{A/B}=\frac{1}{Z} e^{-H^{A/B}}
\label{entanglementH}
\end{equation}
are characterized by free-fermion {\em entanglement} Hamiltonians $H^{A/B}$ ($Z$ is determined by $tr \rho^{A/B}=1$). Eigenvalues and modes of $H^{A/B}$ can be obtained separately\cite{correl}. Here we follow use the method of Klich\cite{klich} and obtain them simultaneously: Introducing an $N_F \times N_F$ Hermitian matrix, $M_{kk'}=\braket{P_A k'}{P_A k}$, where $P_{A(B)}$ are projection operator into subspace $A(B)$ and $\ket{k}$'s are the lowest $N_F$ eigenvectors of original Hamiltonian. Diagonalizing $M$ to obtain its eigenvalues $\zeta_l$ and corresponding eigenmodes $\ket{l}_{Klich}$, the corresponding eigenvalues and (un-normalized) eigenmodess of $H^{A/B}$ can be obtained from
\begin{eqnarray}
\zeta&=&\frac{1}{1+e^{\epsilon_A}}=\frac{1}{1+e^{-\epsilon_B}};\\
|l\rangle_{A/B}&=&P_{A/B}\ket{l}_{Klich}.
\end{eqnarray}
EE takes the form
\begin{equation}
\text{EE}=-\sum_{l=1} ^{N_F} [\zeta_l \ln(\zeta_l)+(1-\zeta_l) \ln(1-\zeta_l)].
\end{equation}
The Klich mode with $\zeta$ closest to $1/2$ (or equivalently $\epsilon_{A/B}$ closest to $0$) has maximum contribution to EE and we call it maximally entangled mode, $\ket{MEM}$. Once projected to subsystem A or B, it is the lowest-energy mode of the corresponding {\em entanglement} Hamiltonian. Its counterpart for the {\em original} Hamiltonian is the energy eigenmode at Fermi energy, denoted as $\ket{E_F}$. We will  make detailed comparisons between $\ket{MEM}$ and $\ket{E_F}$.

To quantify the spatial extent of a mode, we use inverse participation ratio (IPR):
\begin{equation}
\text{IPR}=\frac{1}{\sum_i | \psi_i |^4}.
\end{equation}
where  bigger IPRs corresponds to more extended states.

{\em  Models and Entanglement Entropy} ---
The first model we study here is random dimer model (RDM)\cite{dunalp,note} which is a one dimensional tight binding model with on-site energies $\phi_n$ and constant nearest neighbor
coupling $t$ (we use open boundary condition):
\begin{equation}\label{oh}
H=t\sum_{n=1}^{N-1} (c_n^{\dagger} c_{n+1}+c_{n+1}^{\dagger} c_n)+ \sum_{n=1}^{N} \phi_n c_n^{\dagger} c_n,
\end{equation}
where $\phi_n$ is one of  two independent on site energies $\phi_a$ or $\phi_b$. One of the site energies (here $\phi_b$) is assigned to
neighbouring pairs of lattice sites.  As shown by Dunalp {\em et al.}\cite{dunalp}, when $-2t \leq \phi_a-\phi_b \leq 2t$, states at the resonant energy ($E_{res}=\phi_b$) are
delocalized due to absence of back-scattering. Here we set $t=1$ and $\phi_a=0$, thus this condition reads as: $-2 \leq \phi_b \leq 2$. The system is metallic when the Fermi energy $E_F=E_{res}$, and is insulating otherwise.

\begin{figure}[H]
\centering
\includegraphics[trim = 2cm  7cm 0 0 clip,width=\linewidth]{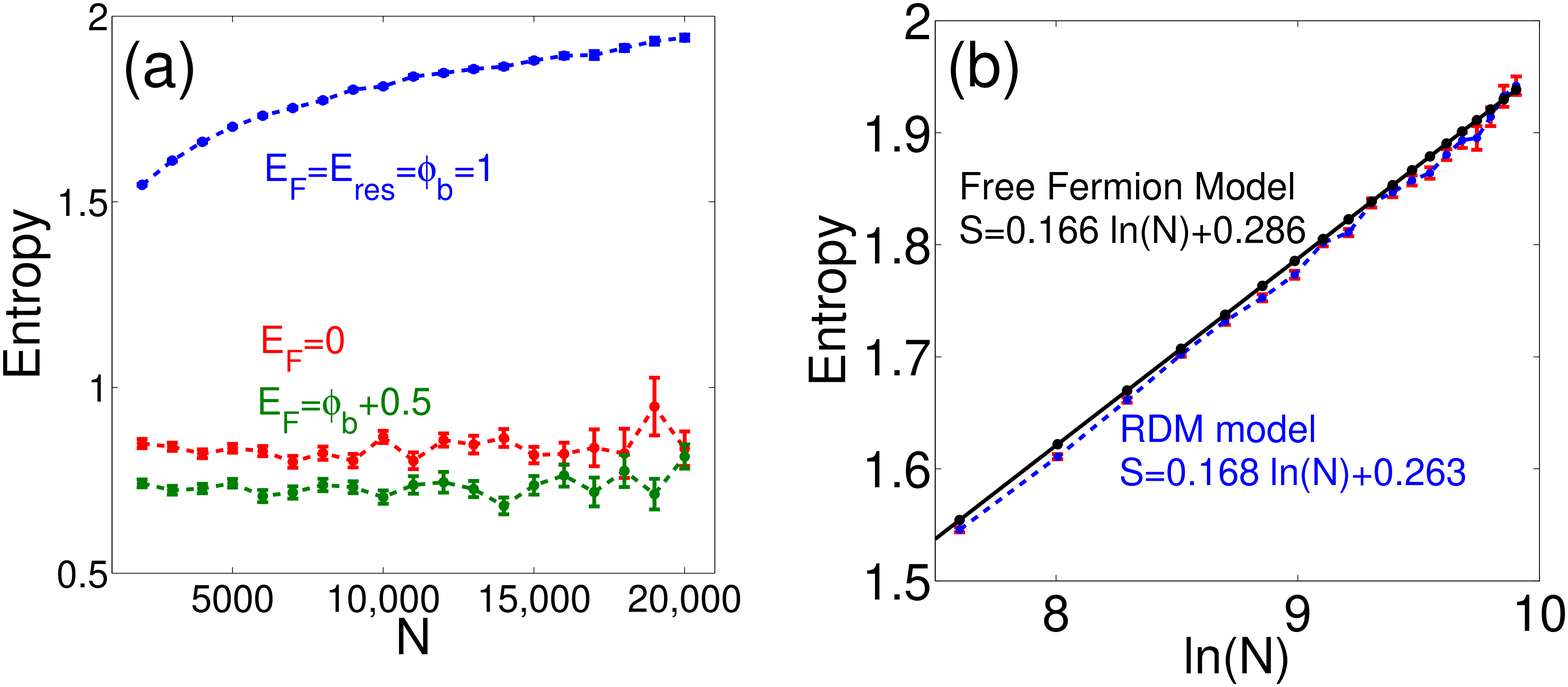}
 \caption{\label{fig:entropyRDM}
(color online) Panel (a): Disorder-averaged entanglement entropy (EE) vs. system size ($N$) for random dimer model for three different $E_F$'s. Subsystem $A$ is from site $1$ to site $N_A=N/2$. Red line: $E_F=0 < E_{res}$, blue line: $E_F=E_{res}$, green line: $E_F=E_{res} +0.5$.  At each point standard error is also included. Number of samples are: $200$ for $N=2000-5000$, $100$ for $N=6000-10000$, $50$ for $N=11000 -15000$, and $20$ for $N=16000-20000$. Panel (b): EE vs. $\ln(N)$ for $E_F=E_{res}$. Blue line is a fitted line for random dimer model entanglement entropy with equation: $EE = 0.168 \ln(N) + 0.263$. Black line is entanglement entropy for free Fermion model which is fitted with line: $EE = 0.166 \ln(N) + 0.266$.}
\end{figure}

We divide the system into two parts: part $A$ from site $1$ to site $N_A=N/2$, and part $B$ from site $N_A+1$ to site $N$. To see how block EE in different phases changes with system
size, we fix $\phi_b=1$, and calculate EE for three different Fermi energies: $E_F=0$, $E_F=E_{res}$ , and $E_F >E_{res}$. The result is plotted in Fig. \ref{fig:entropyRDM}(a). This figure
shows that EE saturates when $E_F=0 < E_{res}$ or $E_F > E_{res}$ with increasing $N$, but when $E_F=E_{res}$, EE diverges, indicating violation of area law. In Fig.
\ref{fig:entropyRDM}(b), EE (when $E_F=E_{res}$) is plotted as a function of $\ln (N)$ and found to follow a straight line. In this plot we also include EE of free fermion hopping model [Eq.
(\ref{oh}) with $\phi_n=0$; we choose $E_F=0$] where we know EE follows a straight line in log-linear scale as $EE = 1/6 \ln(N) + {\rm const}.$\cite{peschel}. We see that EE of both RDM model (when we set $E_F=E_{res}$) and free fermion hopping model have logarithmic dependence on sub-system size, with the {\em same} coefficient. We note this is a {\em stronger} resemblance than that between pure and random spin\cite{refael} or anyon\cite{bonesteelyang} chains, where the coefficient of the log is different.

Another model we study is power-law random banded model (PRBM)\cite{prbm} which is a one dimensional long range hopping model. In this model Hamiltonian matrix ($h_{ij}$) is a random real symmetric matrix with elements following Gaussian distribution with zero mean and variance (we use periodic boundary condition):
\begin{equation}
\avg {\abs{h_{ij}}^2} = \left[{1+\left(\frac{\sin{\pi (i-j)/N}}{b \pi /N}\right)^{2\alpha}}\right]^{-1}.
\end{equation}
At $\alpha=1$ it undergoes Anderson transition (delocalized phase for  $\alpha < 1$ and localized phase for  $\alpha>1$) for every $b$ (we use $b=1$ in this paper).

\begin{figure}
\includegraphics[trim = 2cm  5cm 0 0 clip,width=\linewidth]{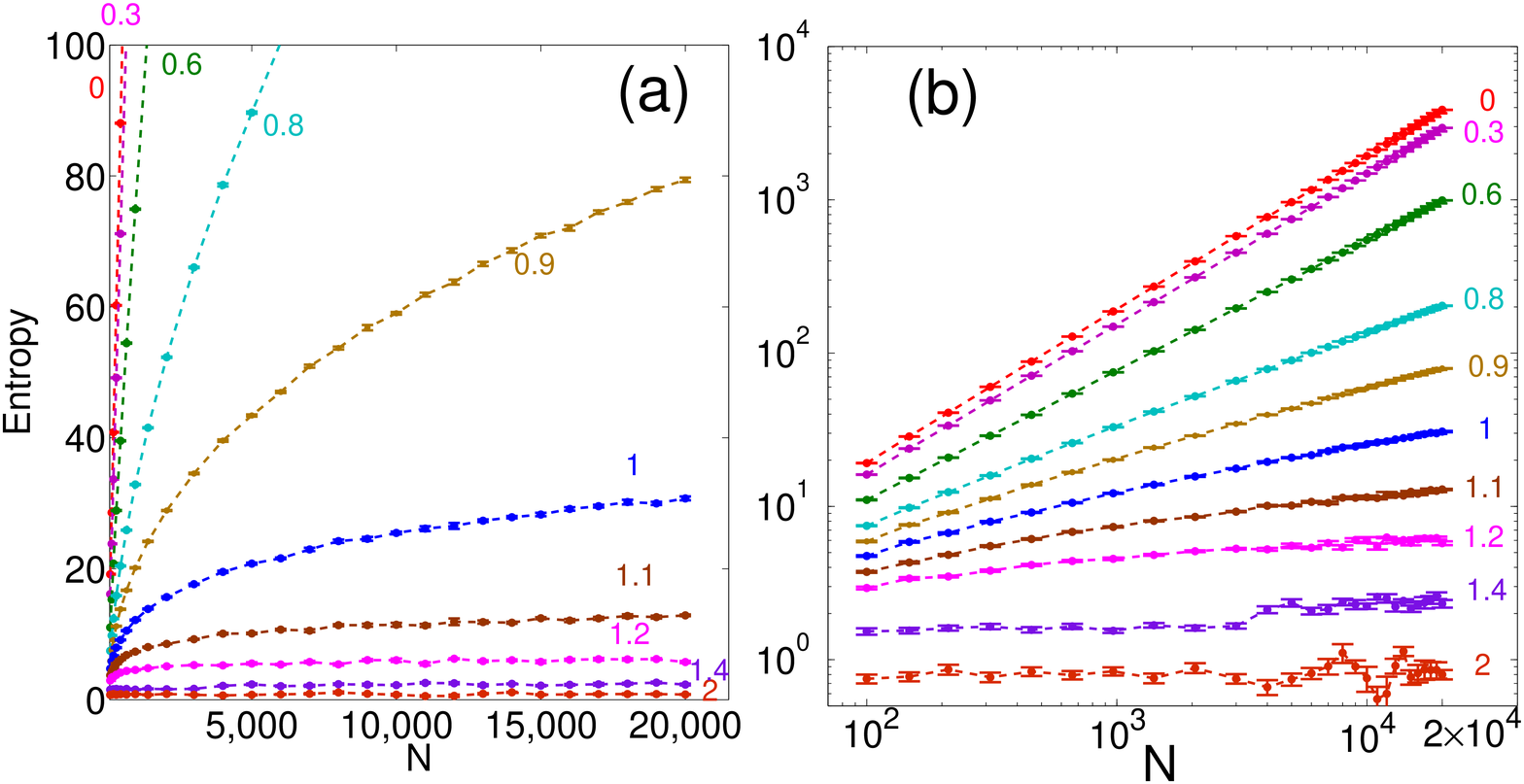}
\caption{\label{fig:PRBMentropy} (color online)  Panel (a): Disorder-averaged entanglement Entropy for power-law random banded model for subsystem $A$ when we divide the system into equal parts as number of sites ($N$) changes, and for different $\alpha$'s. Number of sites changes from $100-20000$. Number of samples are: $100$ for $N=100-5000$, $50$ for $N=6000-10000$, and $10$ for $N=11000-20000$. Panel (b): same data in log-log scale.}
\end{figure}

We calculate EE as a function of system size for different values of $\alpha$, and fix $E_F=0$. Here system is divided into two equal subsystems. The result is plotted in Fig. \ref{fig:PRBMentropy}. We see for $\alpha$'s bigger than $1$ EE saturates, but for $\alpha$'s smaller than $1$, i.e. in metallic phase, EE diverges [Fig. \ref{fig:PRBMentropy}(a)] with increasing system size. We find on log-log scale [Fig. \ref{fig:PRBMentropy}(b)], EE for each $\alpha \le 1$ can be fit with a straight line with equation of $\log_{10} EE=m \log_{10} N + b$. Thus, EE obeys a power-law dependence on system size. Table \ref{table:alpham} shows values of $m$. Such power-law violation of area law is {\em stronger} than previously known examples both in 1D and higher dimensions, which are only logarithmic. It is due to the presence of long-range couplings in this model. In particular, for the case of $\alpha=0$, the hopping strength is {\em independent} of distance and we are thus dealing with an infinite-range model; it is perhaps not surprising that we find $EE\propto N$, namely entanglement entropy follows a volume law instead\cite{note1}.

\begin{table}
\caption{\label{table:alpham}Slope, $m$, of fitted line of power-law random banded model entanglement entropy in the log-log scale. Standard errors of $m$ are beyond written digits. }
\begin{tabular}{l|llllll}
$\alpha$ & 0 & 0.3 & 0.6 & 0.8 & 0.9 & 1 \\
\hline
$m$  & $1.00 $& $0.98$ & $0.84$ & $0.61$ & $ 0.47$ & $0.33 $\\
\end{tabular}
\end{table}

{\em Maximally Entangled Modes and States at Fermi Energy}
---
Li and Haldane\cite{lihaldane} pointed out that the reduced density matrix contains much more information than the entanglement entropy, and showed that spectrum of the entanglement Hamiltonian resembles edge Hamiltonian of a topological state. In Ref. \onlinecite{xxchain} we went one step further by showing that low-lying eigen {\em modes} of the (free-fermion) entanglement Hamiltonian of random XX spin chain contains highly non-trivial information about the system. Here we apply this idea to the present models, and compare the maximally entangled mode $\ket{MEM}$, which is the lowest energy entangled mode of entanglement Hamiltonian and contributes most to entanglement, with the Hamiltonian eigenstate at the Fermi energy $\ket{E_F}$, which contributes most to transport.

\begin{figure}
\includegraphics[width=\linewidth]{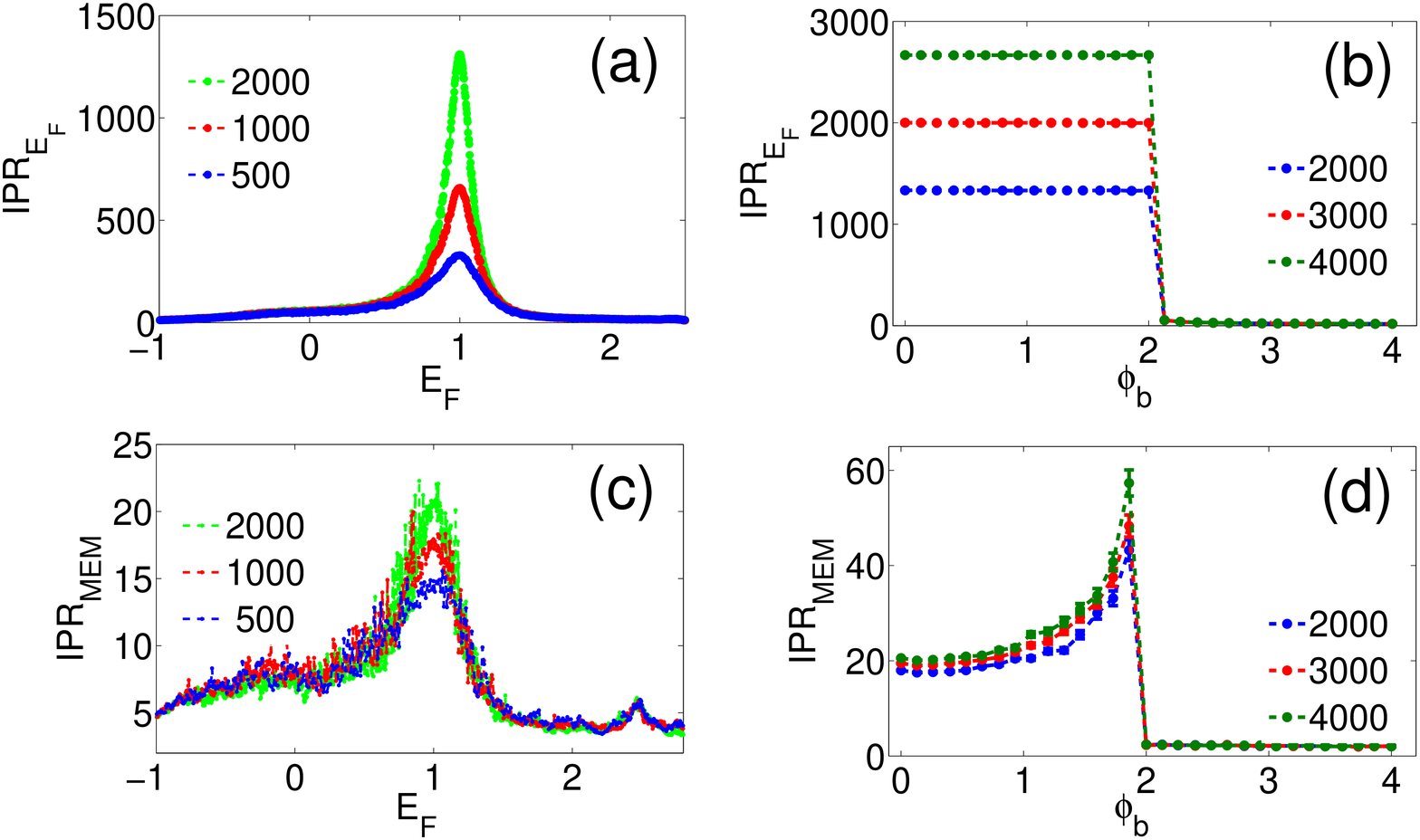}
\caption{\label{fig:RDMIPRs}(color online) Panel (a): Disorder-averaged inverse participation ratio (IPR) of $N_F$th eigenmode of original Hamiltonian for random dimer model as a function of Fermi energy $E_F$. We set $\phi_b=1=E_{res}$. Average is over $100$ samples. Panel (b): IPR of $N_F$th eigenmode of original Hamiltonian for random dimer model as a function $\phi_b$, with $E_F=\phi_b$. Metal-insulator transition occurs at $\phi_b=2$. Average is over $200$ samples. Panel (c): IPR of maximally entangled mode as a function of $E_F$. $\phi_b=1$ as in panel (a). Average is over 100 samples. Panel (d): IPR of maximally entangled mode as a function of $\phi_b$. $E_F=\phi_b$ as in panel (b). Average is over $200$ samples.}
\end{figure}

\begin{figure}
\includegraphics[trim = 0  12cm 0 0 clip, width=\linewidth]{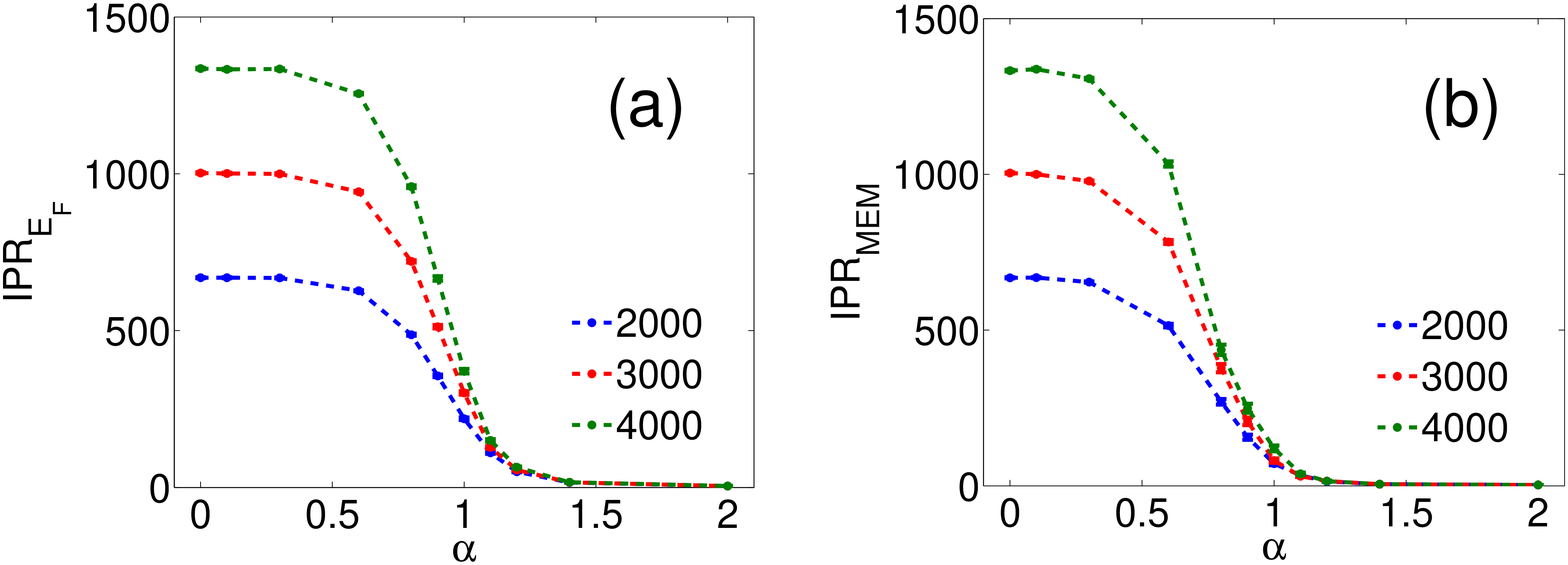}
\caption{\label{fig:PRBMIPRs}(color online)  Panel (a): Disorder-averaged IPR of $N_F$th eigenmode of power-law random banded model Hamiltonian as a function of $\alpha$. $E_F=0$ at each point of $\alpha$. Average is over $200$ samples. Panel (b): IPR of  maximally entangled mode of entanglement Hamiltonian. $E_F=0$ as a function of $\alpha$. Average is over $200$ samples.}
\end{figure}

We start with the RDM. As Fig. \ref{fig:RDMIPRs}(a) shows, IPR of $\ket{E_F}$ has a sharp peak at $E_F=E_{res}=\phi_b$, where the state is delocalized due to absence of back scattering.
Very similar behavior is seen for $\ket{MEM}$, as shown in Fig. \ref{fig:RDMIPRs}(c).
Alternatively, we can also set $E_F=\phi_b$ and observe how $\ket{E_F}$ evolves as a function of $\phi_b$. Fig. \ref{fig:RDMIPRs}(b) shows the result: $\ket{E_F}$ is extended in delocalized phase ($-2\leq \phi_b \leq 2$) and has IPR close to system size $N$, but
they have small IPR in localized phase.
A more interesting behavior is observed for $\ket{MEM}$ in this setting as shown in Fig. \ref{fig:RDMIPRs}(d): As the metal-insulator critical point is approached from the metallic side, its IPR gets {\em enhanced}, and the reduces quickly once entering the insulating phase. Very similar behavior is found in PRBM. As shown in Fig. \ref{fig:PRBMIPRs},
disorder-averaged IPR of $\ket{E_F}$ and $\ket{MEM}$ are almost identical in this case.

\begin{figure}
\includegraphics[trim = 2cm  13cm 0 0 clip, width=\linewidth]{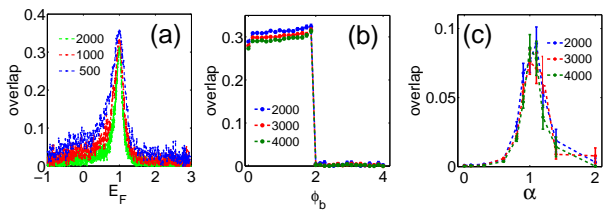}
\caption{\label{fig:overlap}(color online) Panel (a): Disorder-averaged overlap between maximally entangled mode and $N_F$th eigenmode of random dimer model Hamiltonian as we change the Fermi energy. $\phi_b=1$. Average is over $100$ samples. Panel (b): overlap between maximally entangled mode and $N_F$th eigenmode of original Hamiltonian as we change $\phi_b$. $E_F=\phi_b$ at each point of $\phi_b$. Average is over $200$ samples. Panel (c): overlap between maximally entangled mode and $N_F$th eigenmode of original Hamiltonian for power-law random banded model as we change $\alpha$. $E_F=0$ as a function of $\alpha$. Average is over $100$ samples.}
\end{figure}

We can also compare $\ket{E_F}$ and $\ket{MEM}$ more directly by calculating the overlap between them,
$\abs{\braket{E_F}{MEM}}^2$. Their overlap in RDM model is plotted in Fig. \ref{fig:overlap}(a) as a function of Fermi energy. We see a sharp peak at $E_F=E_{res}$, where the system turns metallic from being insulating. Fig. \ref{fig:overlap}(b) shows the overlap as a function of $\phi_b$, with Fermi energy $E_F=\phi_b$. The overlap is much bigger for $\phi_b < 2$ (metallic) than $\phi_b > 2$ (insulating), and again peaks at $\phi_b = 2$ where the transition occurs. In PRBM model, we again see this overlap peaks (quite sharply) at metal-insulator transition ($\alpha=1$), as shown in Fig. \ref{fig:overlap}(c).
We note the overlap at the critical points are quite substantial in both models, given the large systems sizes.

{\em Discussion and Outlook} --- In clean systems, it is well known that free fermion ground states with Fermi surfaces/points violate the entanglement area law. In fact they were the only known examples of area-law violation above 1D\cite{GioevKlich} until the recent discovery that Bose metals with Bose surfaces violates area-law in a similar logarithmic fasion\cite{lai}. Fermi surfaces/points, of course, do not exist in the presence of disorder. However our results suggest that they are {\em not} essential for area-law violation; as long as states at the Fermi level are extended, or equivalently, the system is metallic, such violation occurs. As discussed in the Introduction, dynamic information (in addition to the single Slater determinant ground state) was needed to distinguish between metallic and insulating phases. Our results suggest that by inspecting the entanglement entropy of the ground state {\em alone}, we may be able to distinguish metallic and insulating phases. Furthermore we find certain dynamical property may be {\em extracted} from the ground state entanglement properties; for example the maximally entangled mode, which is the lowest energy eigenmode of the {\em entanglement} Hamiltonian, resembles energy eigenmode at the Fermi energy in intriguing ways, especially near the metal-insulator transition. Metallic phase is, of course, not generic in 1D. The 1D models studied here have special properties that render metallic behavior possible. Nevertheless we are quite optimistic that the conclusions drawn from their studies are robust, and apply to high-dimensions as well. In fact the reason that the metallic phase is stable in the PRBM model is because long-range hopping increases the effective dimensionality of this model. Study of 3D Anderson model is currently underway\cite{py}.

Distinguishing between metallic and insulating phases is more difficult for interacting fermions, as the notion of single-particle states at Fermi energy does not exist. We conjecture that the area-law violation can be used as diagnostic of metallic phase in this case as well. In 1D this is indeed found to be the case in an Anderson/Hubbard model with sufficiently strong attractive interaction\cite{zhao}. This is very reasonable as disorder is {\em irrelevant} and flows to zero under renormalization group in the metallic phase\cite{gs}, and the logarithmic violation of area law has the same origin as that in disorder-free Luttinger liquid. Also there is {\em finite} pairing interaction at the (disorder-free) fixed point theory describing the critical point, thus the (interaction-driven) transition studied in Ref. \onlinecite{zhao} is more appropriately characterized as a superconductor (or superfluid)-insulator transition in 1D. In higher dimensions violation of area law is {\em not} expected for superconductors or superfluids. On the other hand in higher dimensions it was found that Fermi liquid interaction does not alter the logarithmic violation of area law in the disorder-free case\cite{dingprx12}. We expect this violation to be more general however, including in cases where disorder does {\em not} flow to zero as in the metallic phase of 3D Anderson/Hubbard model as well as in the PRBM model studied here.

For an interacting ground state, the entanglement Hamiltonian is no longer free, and the (single fermion) entangled modes cannot be defined. We note it was shown very recently\cite{grover} that the reduced density matrix can be expressed as a sum of terms of the form of Eq. (\ref{entanglementH}), each with a free fermion Hamiltonian in the exponential. It thus may be possible to extract an analog of the maximally entangled mode in some cases.

We close by noting that entanglement properties of {\em single particle} energy eigenstates have been studied in the past\citep{spe}, which is different from our work here which focuses on entanglement properties of the {\em many-body ground state}. Also, some works have been done on characterizing phase transition by studying the \emph{spectrum} of entanglement Hamiltonian (in real and momentum space)\cite{spec}.

\acknowledgements
One of us (KY) thanks David Huse for a useful conversation. This research is supported by the National Science Foundation through grant No. DMR-1004545.

\end{document}